\documentclass[letter,nofootinbib,12pt,superscriptaddress]{revtex4-2}

\usepackage[colorlinks,citecolor=blue]{hyperref}
\usepackage{amsmath}
\usepackage{mathrsfs}
\usepackage{bbm}
\usepackage{amsfonts}
\usepackage{amssymb}
\usepackage{latexsym}
\usepackage{graphicx}
\usepackage[english]{babel}
\usepackage{multirow}
\usepackage{float}
\usepackage{url}
\usepackage{slashed}
\usepackage[compat=1.1.0]{tikz-feynman}
\usepackage{orcidlink}

\begin{document}

\title{A Review of Axion Lasing in Astrophysics}
 
\author{Liang Chen
\orcidlink{0000-0002-0224-7598} \, }
\email{bqipd@protonmail.com}
\affiliation{School of Fundamental Physics and Mathematical Sciences,
Hangzhou Institute for Advanced Study, UCAS, Hangzhou 310024, China}
\affiliation{University of Chinese Academy of Sciences, 100190 Beijing, China}

\author{Thomas W. Kephart
\orcidlink{0000-0001-6414-9590} \, }
\email{tom.kephart@gmail.com}
\affiliation{Department of Physics and Astronomy, Vanderbilt
University, Nashville, TN 37235, USA}

\date{\today}

\begin{abstract}
Axions can be stimulated to decay to photons by ambient photons of the right frequency or by photons from decay of neighboring axions. If the axion density is high enough the photon intensity can be amplified, which is a type of lasing or an axion maser. Here we review the  astrophysical situations where axion lasing can appear and possibly be detected.
\end{abstract}

\maketitle

\section{Introduction}
The Peccei-Quinn mechanism \cite{Peccei:1977hh} is of particular interest because it is still viewed as the most credible solution to the strong CP problem.  The spontaneous breaking of Peccei-Quinn symmetry implies the existence of the spin-0 pseudo-Goldstone boson referred as the axion \cite{Weinberg:1977ma},  which then becomes a candidate for the dark matter. Effective field theory reduces the coupling between axion and Standard-Model particles to
\begin{flalign}
{\alpha_s \over 8\pi f_a}a G^a_{\mu\nu} \tilde G^{a\mu\nu} + {\alpha K \over 8\pi f_a}a F_{\mu\nu} \tilde F^{\mu\nu} + {1\over f_a} J^\mu \partial_\mu a~,
\end{flalign} 
where $\alpha$ and $\alpha_s$ are fine structure coefficients of electromagnetic and strong interactions respectively; $a$, $F_{\mu\nu}$ and $G_{\mu\nu}$ are fields of the axion, QED and QCD respectively; $\tilde F_{\mu\nu}$ and $\tilde G_{\mu\nu}$ and the corresponding dual fields of QED and QCD respectively; $J^\mu$ is the Noether current of the broken symmetry, $K$ is a model dependent coefficient and the decay constant $f_a$ is the energy scale that suppresses the coupling of axions to Standard Model particles. Axions with $f_a$ of electroweak symmetry breaking scale $v_\text{weak}\sim250$ GeV were ruled out \cite{Barshay:1981ky, Barroso:1981ta} by experiments, which left models namely KSVZ \cite{Kim:1979if, Shifman:1979if} and DFSZ \cite{Dine:1981rt, Zhitnitsky:1980tq} predicting invisible axions with $f_a\gg v_\text{weak}$. Also, effective field theory \cite{GrillidiCortona:2015jxo} determines the product of axion mass $m_a$ and decay constant $f_a$ via
$
m_a   f_a = m_\pi f_\pi = [75.5\text{ MeV}]^2 ~,
$
where $m_\pi$ and $f_\pi$ are the mass and decay constant of the pion, respectively.

The axion couples to photons via the interaction
$
{\alpha K \over 8\pi f_a}a F_{\mu\nu} \tilde F^{\mu\nu}, 
$
which suggests the axion lifetime is
\begin{flalign}\label{lifetime}
\tau_a = {256\pi^3 \over K^2\alpha^2 m_a} \bigg({  f_a \over m_a }\bigg)^2
\end{flalign} 
where model dependent coefficient $K$ could be on the order of 1 to 10 \cite{Cheng:1995fd}. The lifetime \eqref{lifetime} of the axion could exceed the age of universe if $m_a\lesssim$ a few eV. This coupling also gained astrophysical interest because of axion-photon conversion in external magnetic fields \cite{Sikivie:1983ip},  could become a strategy for detecting axions. Extensive reviews and references to most of the early work on axions in astrophysics and cosmology can be found in \cite{Kim:1986ax,Raffelt:1990yz,Marsh:2015xka,Braaten:2019knj,Sikivie:2020zpn}.

Axions can be stimulated to decay to two photons by ambient photons of the right frequency or by photons from the decay of other neighboring axions.
Formation of axion miniclusters and Bose stars from axion self-interaction or gravity are discussed in \cite{Kolb:1993zz, Kolb:1993hw, Braaten:2015eeu}. If the axion density is high enough the photon intensity can be amplified, which is a type of lasing or masing. There are several astrophysical situations where axion lasing  appears possible. These include various dense dark matter axion clusters \cite{Tkachev:1986tr,Kephart:1986vc} configurations and axions produced by superradience around primoidial black holes \cite{Rosa:2017ury}. We review and summarize the current possibility of detecting  these astrophysical lasers.

\section{Axion masers}

\subsection{Maser luminosity}

Tkachev  \cite{Tkachev:1986tr} examined the possibility that the growing axion density in the center of a gravitational well gives rise to a coherent cosmic maser source through the stimulated $a\rightarrow\gamma\gamma$ process, during the galaxy formation era. For one model, upon collapse of the irregularities in the axion medium, a substantial luminosity develops: $L \sim m_a^3 \Delta p~ r_\text{p}^2 \exp(D)$ where $r_\text{p}$ is a gravity related distance, $\Delta p$ is the uncertainty in axion momentum, and $D$ is the integrated amplification factor, as defined below. The luminosity $L$ is comparable with that of the brightest quasars provided $D\sim10^2$, which is attainable if $M<10^8 [(10^{12} \text{GeV})/f_a]^4 M_\odot$, where $M$ is the mass of irregularities  and $M_\odot$ is the solar mass.

Assuming that cosmological strings \cite{Kibble:1976sj} are the seeds of gravitational condensation, and considering the formation of spherically symmetric axionic configuration due to the velocity dispersion, phase space density(occupation number) $f_a$ is isotropic within the ``core'' region \cite{Tkachev:1987cd}. If the photon phase space density is far less than the axion phase space density($f_\gamma\ll f_a$) and the gravitational field is weak, then for a wide class of axion distribution $f_a$, the photon phase space density
$f_\gamma$ evolves according to
\begin{flalign}
{df_\gamma\over dt} =
\frac{\pi^2\Gamma_a}{m_a^4 \beta} \epsilon (1+  f_\gamma )  
\end{flalign}
where
\begin{flalign}
\epsilon = m_a \int f_a(\vec{p},\vec{r}) {d^3 p \over (2\pi)^3}
\end{flalign}
is the axion energy density, $\beta$ is the axion escape velocity characterizing the depth of the gravitational well.
$f_\gamma$ increases exponentially with time if $D>1$ is satisfied, where the amplification coefficient $D$ is introduced as
\begin{flalign}
D(r)=\int_0^r {d\over dt}f_\gamma(r') dr'~.
\end{flalign}
For a self-gravitating system, the amplification becomes \cite{Tkachev:1987cd}
\begin{flalign}\label{ec}
D\sim {f_\pi\over r_c f_a}~,
\end{flalign}
where $r_c$ is the radius of the core region.
$D>1$ gives \cite{Tkachev:1987cd}
\begin{flalign}\label{rcless}
r_c<10^7 \beta {10^{12}\text{ GeV}\over f_a}\text{ cm}~.
\end{flalign}

If the characteristic energy scale of the symmetry breaking during phase transitions in the early Universe is $\sqrt \mu$, then the linear mass density of cosmic strings is $\mu$, and the network of strings produces oscillating loops of size $r_s\sim ct$ at time $t$.
A spherical shell of radius $r_s$ and mass $2\pi r_s\mu$ may approximate gravitational effect of the string loop \cite{Sato:1986csa}. If axions initially were
inside the loop and form the core, then they can constitute galactic halos \cite{Sato:1986csa} and maintain the energetics of the core. A rapidly moving string alters axion trajectories which suggests that if $r_c$ is equal to the loop size $r_s$, this may violate stimulated emission condition \eqref{rcless}. But the effect of higher multipole moments is small in   distant regions \cite{Silk:1984xk}, and axions are not disturbed by string motion and if the core formed by axions can reach a smaller radius, then the enhancement coefficient \eqref{ec} is still a valid choice. Thus, this region can provide a luminosity of \cite{Tkachev:1987cd}
\begin{flalign}\label{}
L_0 \sim 10^{-15} M_\text{pl}^2 {m_s\over M_\odot} \left({\text{1 yr}\over t_c}\right)^{1\over3}~,
\end{flalign}
where $t_c$ is the recollapse time for the axions and $ M_\text{pl}$ is the Planck mass.

Axion accretion on strings may result in a core with a very high stimulated emission rate, or a quasar-like energy release. A pure axionic object produces a monochromatic spectrum with frequency centered at $m_a/2$. To produce the complicated spectrum of quasars, the axionic core in a gravitational well may be surrounded by ordinary(baryonic) matter \cite{Tkachev:1986tr}.
If one cannot identify certain lines in a spectra with any molecular or atomic level, then the axion mass could be determined \cite{Tkachev:1987cd}.
Odd-integer harmonics of the fundamental
frequency could also be produced by interactions $3a\rightarrow2\gamma$ at much lower magnitude \cite{Braaten:2016dlp}.

Tkachev \cite{Tkachev:2014dpa} proposes that the properties (energy release, duration, event rate) of FRB can be matched with the explosive maser effect of axion miniclusters or the decay of axions in external magnetic field.

\subsection{Parametric resonance}

Levkov et al. \cite{Levkov:2020txo} have developed a quasistationary formalism of parametric resonance in a finite volume for nonrelativistic axions, incorporating (in)coherence, finite-volume effects, axion velocities, binding energy, gravitational redshift, and backreaction of photons on axions. Axions with large occupation numbers are described by a classical field $a(t,\vec{x})={f_a\over\sqrt2}[\psi(t,\vec{x}) e^{-im_a t} + \text{h.c.}]$ and affected by a potential
\begin{flalign}\label{}
V={m_a^2\over2}a^2 - {g_4^2 m_a^2 \over4!f_a^2}a^4 + ... \  ~.
\end{flalign}
If $\lambda_a$ is the wavelength of the axion, then the nonrelativistic approximation
reads
\begin{flalign}\label{}
m_a\lambda_a \gg1~,~
\partial_t\psi\sim\psi/(m_a\lambda_a^2)~,~
\partial_{\vec x}\sim\psi/\lambda_a ~.
\end{flalign}
A stationary solution of the Schrödinger-Poisson system gives a Bose-Einstein condensate of axions in the ground state of a nonrelativistic gravitational potential $\Phi$, for
$\psi = e^{-i\omega_s t} \psi_s(r)$, where $\omega_s<0$ is the binding energy of axions. Since the phase of $\psi_s$ is independent of spacetime, the axions are coherent \cite{Levkov:2020txo}.
Consider the electromagnetic potential $A$ along an arbitrary $z$-direction,
\begin{flalign}\label{}
A_i=C_i^+ e^{im_a(z+t)/2} + 
C_i^- e^{im_a(z-t)/2} + \text{h.c.} ~.
\end{flalign}
Substituting $A_i$ into Maxwell’s equations shows that the electromagnetic field changes fast, $\partial_t C\sim C/\lambda$, which prompts adiabatic the ansatz \cite{Levkov:2020txo},
\begin{flalign}\label{}
C_i^\pm = e^{\int^t \mu(t')dt'} c_i^\pm(t,\Vec{x}) ~,
\end{flalign}
where $\mu$ and the quasistationary amplitudes and $c_i^\pm$
evolve on the same time scales $m_a\lambda^2$ as $\psi$.
Static homogeneous axions in an infinite volume give $\partial_z c_i^\pm=0$ and $\mu_\infty=g m_a|\psi|$, where $g \sim \alpha K  $ is the dimensionless axion-photon coupling. Let $L$ be the size of an axion cloud; photons accumulate if $\mu_\infty L \gtrsim1$.

Assuming that $\psi$ is real, which means the axions are static and coherent, there  exist localized solutions with
\begin{flalign}\label{}
c_i^+|_{z\rightarrow+\infty} = 
c_i^-|_{z\rightarrow-\infty}=0 ~,
\text{ Re }\mu\geq0 ~,
\end{flalign}
representing resonance instabilities. Defining $D(z)$ as
\begin{flalign}\label{ResoCondi}
D(z) = g m_a \int_{-\infty}^z dz' \psi(z')
~,
\end{flalign}
parametric resonance along the arbitrary $z$-direction corresponds to $D(+\infty)\geq{\pi\over2}$. Resonance starts with a small exponent $\mu\ll L^{-1}$ immediately after the condition \eqref{ResoCondi} is met which means initial growth is tiny. When the electromagnetic amplitude become large, the backreaction will cause the resonant flux to fall immediately, but a long-lived quasistationary  electromagnetic field could appear, causing a glowing axion star to be formed \cite{Levkov:2020txo}. 

Levkov et al. \cite{Levkov:2020txo} argue that the Bose stars with $D(+\infty)\ll1$ are better amplifiers(stimulated decay) than diffuse axions, when an external  radio wave of frequency $\sim m_a/2$ travels through the axions. 
For diffuse axion cloud, photon fluxes could be amplified but exponential growth is not expected.
The case of collapsing axion stars is also investigated in \cite{Levkov:2020txo}: the star initially contracts without electromagnetic effect, growth of the luminosity begins once the localized solution appears.

Other notable findings  in \cite{Levkov:2020txo} include: the modes with different angular harmonic number $l$ of a spherical axion star grow at rate $\sim\mu$,
resonance may develop when two axion stars come close to each other with negligible relative velocity even if resonance does not occur for individual Bose stars, and as expected
relative velocities among axions prohibit resonance, etc.

Another important type of astrophysical axion objects are axion bose condensates found in the work of Sikivie, et al.
\cite{Sikivie:2009qn,Erken:2011dz}
and further analysed by
Hertzberg and Schiappacasse \cite{Hertzberg:2018zte}, who focused on axion clump resonance of photons. They correspond to unstable (resonant) and stable solutions of the Mathieu equation, which is the equation of motion for homogeneous small amplitude axion fields.  The electromagnetic modes have a maximum exponential growth rate of \cite{Hertzberg:2018zte}
\begin{flalign}\label{}
\mu_H^* \approx {\alpha K \over 8\pi f_a} m_a a_0  ~,
\end{flalign}
in the first resonant region, where $a_0$ is the amplitude of homogeneous
axion oscillation $a(t)$. Note that $\mu_H^*$ in \cite{Hertzberg:2018zte} is comparable with  $\mu_\infty$ in \cite{Levkov:2020txo}.

The homogeneous axions may eventually become unstable and collapse/condensate towards an axion clump, from gravity and attractive self-interactions. A spherically symmetric axion clump was found \cite{Schiappacasse:2017ham} to be accurately approximated by
\begin{flalign}\label{}
\psi = \sqrt{3N\over\pi^3 R^3} \text{ sech}\left({r\over R}\right) e^{-i\mu t}  ~,
\end{flalign}
where $R$ is the radius of the clump, $\mu$ serves as a correction to the frequency $m_a/\hbar$, and $N=\int d^3x \psi^*\psi$ is the particle number of axions. A clump of $N$ axions cannot resonate if $N<N_c$ where $N_c$ is a critical value of the particle number. In the case of attractive axion self-interactions, there is a maximum particle number $N_\text{max}$ in an axion clump \cite{Hertzberg:2018zte}, 
\begin{flalign}\label{}
N_\text{max} \approx {10.12 f_a \over |g_4| \sqrt{G} m_a^2 }  ~.
\end{flalign}
If a clump  of QCD axions has $N_\text{max}$ and attractive self-interactions and $g_4^2$=0.3 (the preferred value for conventional QCD axions) \cite{GrillidiCortona:2015jxo}, there would be no resonance due to axion-photon coupling coefficient $K$ being too small in current models. (However, resonance of hidden sector photons could still occur \cite{Hertzberg:2018zte}.)  The general criteria for clump resonance is that a pair of photons being produced has to stimulate  another pair of photons before escaping the axion clump, which is succinctly expressed as 
\begin{flalign}\label{}
\mu_H^* \approx {\alpha K \over 8\pi f_a} m_a a_0 >{c\over2R} ~,\,\,\, \text{ where }\,\,\,
a_0= \sqrt{2\over m_a} \sqrt{3N\over\pi^3 R^3} ~,
\end{flalign}
i.e., the homogenous axion field growth rate must be higher than the photon escape rate \cite{Hertzberg:2018zte}. A typical growth time for a spherical clump of QCD axions with $m_a=10^{-5}$ eV is estimated to be $1/\mu^*\lesssim2\times10^{-4}$ s, which is similar to the duration($\sim$ms) of pulses calculated in \cite{Rosa:2017ury}.

A non-spherical clump profile was found to be accurately approximated by a modified Gaussian  \cite{Hertzberg:2018lmt}, which leads to a maximally allowed number $N_\text{max}$ of particles in the clump being larger than that of spherical clump. $N_\text{max}$ of a non-spherical clump increases rapidly with angular momentum $l$ and thus it is easier to achieve resonance of photons. The condition for resonance that $\mu_H^*>{c\over2R}$, carries over from spherical clumps to non-spherical axion configurations. For conventional QCD axions, resonance does not occur in non-spherical clumps unless the angular momentum is very large  $l \gtrsim \mathcal{O}(10^3)$  \cite{Hertzberg:2018zte}. 

A scenario of clump resonance in astrophysics would be: clumps formed under gravity in the past with particle number $N>N_c$ resonate into photons and loss energy individually, which inevitably results in $N<N_c$. Clumps with $N<N_c$ merge together and become a clump with $N>N_c$, which could still resonate today \cite{Hertzberg:2018zte}.

\section{Lasing axions as particles}

\subsection{Spontaneous emission}
Kephart and Weiler \cite{Kephart:1986vc} conservatively proposed that the luminosity of an axion cluster is comparable to that of an astrophysical object (star, galaxy, or galaxy cluster) of similar mass, considering solely the mechanism of spontaneous decay of the axion. Let $N_a\sim10^{66}(M/M_\odot)[(1 \text{ eV})/m_a]$,  be the total number of axions in a cluster of mass $M$, where $L_\odot$ is  the solar luminosity. The luminosity of photons $L$ from spontaneous emission of the cluster is then
\begin{flalign}\label{}
L={m_a N_a\over\tau_a} \sim 4\times10^{29} \left({m_a\over1 \text{ eV}}\right)^5  {M\over M_\odot} {\text{erg}\over \text{s}}
 \sim 10^{-4} \left({m_a\over1 \text{ eV}}\right)^5  {M\over M_\odot} L_\odot  ~.
\end{flalign}
If an axion cluster is at a distance of $D=300 h^{-1} ( z_c/0.1 )$ from the earth, the flux observed at the earth becomes
\begin{flalign}\label{}
F={L\over4\pi D^2} = 4.7\times10^{-16} h^2 f(z_g)  {r\over 1 \text{ pc}} \left({0.1\over z_c}\right)^2  \left({m_a\over1 \text{ eV}}\right)^5  {\text{W}\over \text{m}^2} ~,
\end{flalign}
where $z_c$ is the cosmological redshift, $h$ is related to the present Hubble parameter $H_0$ by $H_0 = 100h$ km/s Mpc, $r$ is the radius of the cluster, and $f(z_g)$ describes the surface gravitational red shift through $f(z_g) \equiv z_g(2+z_g) / (1+z_g)^2   =  2GM/ r$. Suppose that a mass fraction $x_a$ of a galactic halo is axionic, the spectral photon radiance of the Milky Way and the spectral photon irradiance of Andromeda are then found to be
\begin{flalign}\label{}
{dI\over d\Omega \Delta\lambda} \sim 10^4 x_a \left({m_a\over1 \text{ eV}}\right)^5  ( \text{cm}^2 \cdot \text{s} \cdot \text{sr} \cdot \text{\AA} )^{-1}
~,\,\, \text{and} \,\qquad
{I\over \Delta\lambda} \simeq 20 x_a \left({m_a\over1 \text{ eV}}\right)^5  ( \text{cm}^2 \cdot \text{s} \cdot \text{\AA} )^{-1} ~,
\end{flalign}
respectively. Comparing these with the background photons of the relevant spectrum in both galaxies, one can find lower bounds on $x_a$ that allow axions to be detectable.

\subsection{Stimulated emission rate equations}
For any species of particles with occupation number $f(\vec{p},\vec{r},t)$, the particle number density $n(\vec{r},t)$ and the total particle number $N$ in a volume $V$ are, respectively
\begin{flalign}\label{EvoEq01}
n(\vec{r},t)= \int \frac{d^3p}{8\pi^3} ~ f(\vec{p},\vec{r},t), ~\text{and}~~ N = \int_V d^3r ~ n(\vec{r},t).
\end{flalign}
The rate of change of the photon number density $n_{\lambda}$ of helicity $\lambda=\pm 1$ due to the decay process $a\rightarrow\gamma\gamma$ is given by Boltzmann equation,
\begin{flalign}\label{BoltzmannEq}
\frac{dn_{\lambda}}{dt}=\int dX^{(3)}_\text{LIPS}[f_a(1+f_{1\lambda})(1+f_{2\lambda})-f_{1\lambda}f_{2\lambda}(1+f_a)]|M(a\rightarrow \gamma \gamma)|^2 ~,
\end{flalign}
where $f_a(\vec{p})$, $f_{i\lambda}=f_\lambda(\vec{k}_i)$ are occupation numbers of axion and photon respectively. $M(a\rightarrow \gamma \gamma)$ is the decay amplitude of the coupling term ${\alpha K \over 8\pi f_a}a F_{\mu\nu} \tilde F^{\mu\nu}$ and $dX^{(3)}_\text{LIPS}$ is the Lorentz invariant three-body phase space of the axion and two photons.
Assuming spherical symmetry $f_a(\vec{p})= f_a(|\vec{p}|)\equiv f_a(p)$, $f_\lambda(\vec{k})= f_\lambda(|\vec{k}|)\equiv f_\lambda(k)$ for both the axion and photon occupation numbers, one could carry out the integration over the momentum space of one of the photons,
\begin{flalign}\label{EvoEq02}
\frac{dn_\lambda}{dt} =
\frac{m_a \Gamma_a}{2\pi^2} \int_{m_a} d p^0 \int_{k_{\text{\tiny min}}}^{k_{\text{\tiny max}}} dk 
 \{  f_a  ( p^0 ) [ 1+ 2f_\lambda ( k ) ] - f_\lambda ( k )f_\lambda ( p^0-k ) \} ~,
\end{flalign}
where
\begin{flalign}
k_\text{max/min} = { m_a^2 \over 2 ( p^0 \mp p ) } =  \frac{p^0 \pm \sqrt{(p^0)^2-m_a^2}}{2} 
\end{flalign}
is the maximum/minimum photon momentum from the decay of an axion of momentum $\vec{p}$. The rate of change of the axion density is
\begin{flalign}\label{EvoEq03}
\frac{dn_a}{dt} = -{1\over2}\sum_{\lambda=\pm}\frac{dn_\lambda}{dt} ~,
\end{flalign}
which together with Eqs.\eqref{EvoEq01} and \eqref{EvoEq02} become a set of integro-differential equations for the evolution of the axion-photon system \cite{Kephart:1994uy}. For the approximately static clusters to which the formalism has been applied, it is appropriate to treat the axions as particles \cite{Kephart:1999ti}.

\subsection{A simple axion cluster model}
A simple model \cite{Kephart:1994uy} of lasing axion clusters assumes the following forms
\begin{flalign}\label{sss}
f_a(p, r, t)=f_a\Theta(p_{\text{\tiny max}}-p)\, \Theta(R-r) ~,
f_\lambda(k, r, t)=f_\lambda\Theta(R-r)\Theta(k_+-k) \Theta(k-k_-) ~,
\end{flalign}
for axion and photon occupation numbers respectively, where $f_a$ and $f_\lambda$ depend on time only, $R$ is the radius of the cluster, $p_{\text{\tiny max}}=m_a\beta$ is the maximum axion momentum, $\beta$ is the maximum(escape) velocity of an axion and $k_\pm=\frac{m_a\gamma}{2}(1\pm\beta)$ is the maximum/minimum photon momentum from the decay of an axion with the maximum momentum $m_a\beta$. Integration of Eq.\eqref{EvoEq02} over $p$ and $k$ space leads to
\begin{flalign}
\frac{d n_\lambda  }{dt}  = \Gamma_a  \bigg[  n_a \bigg( 1 + \frac{ 16 \pi^2 n_\lambda }{ \beta m_a^3 } \bigg) - \frac{ 32 \pi^2 n_\lambda^2 }{ 3 m_a^3 } \bigg(\beta + \frac{3}{2} \bigg)  \bigg] - {3c n_\lambda \over 2R} ~,
\end{flalign}
where the last term accounts for the surface loss of the photons. The rate of change of the axion number density is half of that of the photon number density without surface losses,
\begin{flalign}
\frac{d n_a  }{dt}  = -{1\over2} \Gamma_a \sum_{\lambda=\pm} \bigg[  n_a \bigg( 1 + \frac{ 16 \pi^2 n_\lambda }{ \beta m_a^3 } \bigg) - \frac{ 32 \pi^2 n_\lambda^2 }{ 3 m_a^3 } \beta  \bigg] ~,
\end{flalign}
where we subtract the rate of production of axions with velocity exceeding $\beta$ from back reaction $\gamma\gamma\rightarrow a$. Since these fast axions escape from the cluster, they are labeled as ``sterile''  with a production rate
\begin{flalign}
\frac{d n_s  }{dt}  = {1\over2} \Gamma_a \sum_{\lambda=\pm}  \frac{ 16\pi^2  }{  m_a^3 } n_\lambda^2   ~.
\end{flalign}
If one assumes a common initial condition $n_+(0) = n_-(0)$, then helicity densities $n_\pm$ evolve identically, and $n_+(t) = n_-(t)={1\over2}n_\gamma(t)$ for all time. Substituting $n_\gamma$ for $n_\pm$ results in equations irrespective of helicity,
\begin{flalign}\nonumber
\frac{d n_\gamma  }{dt}  &= 2 {1\over\tau_a} n_a  + \frac{ 16 \pi^2  }{ \beta m_a^3 \tau_a } n_a n_\gamma - \frac{ 16 \pi^2 }{ 3 m_a^3 \tau_a } \left(\beta + \frac{3}{2} \right) n_\gamma^2    - {3c  \over 2R} n_\gamma ~,\\\label{EvoEq04}
\frac{d n_a  }{dt} &= -\left( {1\over\tau_a} n_a +  \frac{ 8 \pi^2  }{ \beta m_a^3 \tau_a } n_a n_\gamma  -  \frac{ 8 \pi^2 \beta }{ 3 m_a^3 \tau_a } n_\gamma^2  \right) ~,
\quad
\frac{d n_s  }{dt}  = \frac{ 4 \pi^2 }{  m_a^3 \tau_a }   n_\gamma^2 ~.
\end{flalign}
Expressing time and volume in the units of spontaneous axion lifetime $\tau_a$ and Compton volume ${16\pi^2\over m_a^3}$, the set of evolution equations become dimensionless \cite{Kephart:1994uy, Kephart:1999ti},
\begin{flalign}
\dot{\bar n}_\gamma&= 2 \bar n_a  + {1\over\beta} \bar n_a \bar n_\gamma -  \bigg({\beta\over3} + {1\over2} \bigg) \bar n_\gamma^2    - {3c\tau_a\over 2R} \bar  n_\gamma ,~
\dot{\bar n}_a = - \bar n_a - { 1 \over 2\beta }  \bar n_a  \bar n_\gamma  +  {  \beta \over 6 } \bar n_\gamma^2     ,~
\dot{\bar n}_s  = { 1 \over  4 } \bar n_\gamma^2 ~, 
\end{flalign}
where $\bar n_a$, $\bar n_\gamma$ and $\bar n_s$ are dimensionless number densities and the dot derivative is with respect to dimensionless time $t/\tau_a$. The characteristic time for lasing is
\begin{flalign}
\bar\tau_\text{lase} \equiv {\tau_\text{lase}/\tau_a}= \left({\bar n_a(0)\over\beta} - {3c  \over 2R}\right)^{-1}~.
\end{flalign}

\subsection{Parameters and conditions}
The axion cluster has an escape velocity of $\sqrt{2GM/R}$, which is set to be the $\beta$ parameter in the axion lasing model. Axions with higher speed would not be confined by the cluster body of mass $M$,
\begin{flalign}
\beta  = \sqrt{\frac{2GM}{R}} = 1.8\times10^{-3} \sqrt{\frac{\rho_a}{\text{g/cm}^3}} {R\over R_\odot}~,
\end{flalign}
where $\rho_a = {m_a^4 \bar n_a/(16\pi^2)}$ is the mass density of the axion cluster in units of g/cm$^3$, $R_\odot$ is the solar radius. 

An estimate of the necessary values of axion cluster parameters for lasing can be found by requiring the photon mean free path to be less than the diameter of the cluster, which gives
\begin{flalign}
\bigg({m_a\over\text{eV}}\bigg)^2 \, {\rho_a\over\text{g/cm}^3} &\geq \frac{6}{K^4} ~.
\end{flalign}
Another estimation of the values of these parameters for small $t$, neglects the back reaction term in Eq.\eqref{EvoEq04} and demands the coefficients of $n_\gamma$ on the right hand side of the equation be positive,
\begin{flalign}
\frac{16\pi^2 n_a}{\tau_a m_a^3 \beta} - \frac{3c }{2R} >  0 ~.
\end{flalign}
These two lasing conditions reproduce each other up to a factor of 3 and provide a lower bound for the density of the lasing axion cluster. Keeping $\beta\lesssim1/2$ avoids consideration of black hole effects and provides an upper bound for the density of a lasing cluster at a fixed radius. Considering both bounds places the cluster density in the range \cite{Kephart:1994uy, Kephart:1999ti},
\begin{flalign}
{6\over K^4}\bigg({\text{eV}\over m_a}\bigg)^2 \leq {\rho_a\over\text{g/cm}^3}  \leq  7.7\times10^4 \left({R_\odot\over R}\right)^2~.
\end{flalign}
The maximum radius for lasing to occur is obtained by saturating both bounds,
\begin{flalign}
R\lesssim 100K^2 {m_a\over\text{eV}} R_\odot ~,
\end{flalign}
which also leads to the maximum mass of the cluster
\begin{flalign}
M   \leq  6 \times10^6   K^2 {m_a\over\text{eV}}   M_\odot ~,
\end{flalign}
There are no lower bounds for $R$ and $M$.

The eventual lasing is inevitable as it needs only a few photons to trigger lasing, regardless of the initial photon density $\bar n_\gamma(0)$ of the cluster, which regulates the timing of the laser profile. Based on the ratio $R/(c\tau_a)$ between the photon diffusion time $R/c$ and the axion life time $\tau_a$, one can write  the initial photon density from spontaneous emission of an axion cluster as
\begin{flalign}
 \bar n_\gamma(0)=320 \bigg({m_a\over \text{eV}}\bigg) K^2 \beta^2  \bigg({R_\odot\over R}\bigg).
\end{flalign}

Photons cannot propagate if the frequencies are below the plasma frequency (due to high electron density in the early universe), which might inhibit axion decay. It has been shown \cite{Kephart:1994uy} that axion decay proceeds without inhibition as long as the red shift $z\lesssim10^9 (m_a/\text{eV})^{2/3}$.

\subsection{Discussion}
The flat distribution functions in the simple axion cluster model are idealizations. Deviations from a flat distribution will give regions of higher density, which will be more active for lasing. Gravitational collapse could trigger formation of a high axion density in the vicinity of an object such as black hole, where gravitational corrections to the rate equations would have to be included. There are a wide range of collisionless self-gravitating spherically symmetric systems where the density and velocity distribution functions are known in closed analytic form \cite{Fridman:1984Pol}. To obtain general rate equations for realistic distributions by integration is straightforward.

If compact objects with mass larger than $10^7 M_\odot$ in the cores of nearby galaxies were axion clusters, they are not expected to lase since their mass and size fall outside the allowed parameters for lasing. However, spontaneous axion decay may still generate luminous emissions \cite{Kephart:1986vc}. Furthermore, if these massive cluster were made of smaller dense clusters, lasing might still be possible for individual small clusters.

After the Universe became matter dominated, the energy released by detonation of a typical hadronic axion \cite{Kaplan:1985dv} cluster allows evacuation of matter and formation of a void, by multiphoton ionization and the subsequent plasma absorption \cite{Kephart:1994uy}. The total energy available in cosmic axions is sufficient to power the formation of all the voids in the Universe.

\subsection{Application -- Superradiant Clouds}
In Kerr spacetime, the Klein-Gordon equation admits hydrogenic-like solutions localized in the vicinity of black hole(BH), characterized by integer quantum numbers$(n,l,m)$. (For a revied of superradiance see  \cite{Brito:2015oca}.) Let $M_\text{BH}$ and $J_\text{BH}$ be the mass and angular momentum of the black hole respectively. In the nonrelativistic regime, the spectrum of the quasibound state is  \cite{Brito:2015oca}
\begin{flalign}
\hbar \omega_n \simeq m_a c^2\left( 1 - {\alpha_\mu^2\over2n^2} \right) ~, \text{ where } \alpha_\mu \equiv { G m_a M_\text{BH}\over\hbar c } ~.
\end{flalign}
The scalar field extracts the BH’s rotational energy and the axion cloud around the BH grows if $\omega_R<m\Omega_\text{BH}$, where $\omega_R$ is the real part of the complex frequency $\omega=\omega_R+i \omega_I$ and $\Omega_\text{BH}$ is the horizon’s angular velocity. When $\alpha_\mu\ll1$, the occupation
number of the fastest growing mode ``$2p$''$(n=2,l=m=1)$ grows at a rate
\begin{flalign}
\Gamma_s \simeq 4\times10^{-4} \tilde a \left({\mu\over10^{-5}\text{eV}}\right) \left({ \alpha_\mu\over 0.03 }\right)^8 \text{ s}^{-1} ~,
\end{flalign}
where $\tilde a =  c J_\text{BH} / (G M_\text{BH}^2)$ is the BH’s dimensionless spin parameter$(0<\tilde a<1)$. In the nonrelativistic regime the axion cloud is localized far away from the horizon, hence thre formalism developed in \cite{Kephart:1994uy} is still applicable. Although the geometry of the ``$2p$'' state is more intricate than the flat spherically symmetric model, the estimates made by Rosa and Kephart \cite{Rosa:2017ury} by exploiting Eqs.\eqref{EvoEq04} are sufficiently good in terms of the order of magnitude. They calculated the peak luminosity, total energy and duration of of the black hole lasers powered by axion superradiant instabilities(BLASTs). They are
\begin{flalign}
L_{B}&\simeq {2\times 10^{42}\over K^2}\tilde{a} \left({10^{-5}\ \text{eV}\over \mu}\right)^{2}\left(\alpha_\mu\over 0.03\right)^7\left({\xi\over 100}\right)\ \text{erg/s}~,\nonumber\\
E_{B}&\simeq {3\times 10^{39}\over K^2}\sqrt{\tilde{a}} \left({10^{-5}\ \text{eV}\over \mu}\right)^{3}\left(\alpha_\mu\over 0.03\right)^{5\over2}\left({\xi\over 100}\right)^{1\over2}\ \text{erg}~,\nonumber\\
 \tau_{B}&\simeq {1\over \sqrt{\tilde{a}}} \left({10^{-5}\ \text{eV}\over \mu}\right)\left(\alpha_\mu\over 0.03\right)^{-9/2}\left({\xi\over 100}\right)^{-1/2}\ \text{ms}~,\\
\text{where }
\xi&= \log\left({\Gamma_s\over\Gamma_a}\right)\simeq  107-4\log\left({\mu \over 10^{-5}\ \mathrm{eV}}\right)+8\log\left({\alpha_\mu\over 0.03}\right)+\log \left({\tilde a\over K^2}\right) \nonumber ~.
\end{flalign}
Note that the duration $\tau_{B}$ is similar to the growth time for a spherical clump of QCD axions with $m_a=10^{-5}$ eV estimated in \cite{Hertzberg:2018zte}.  The brightest bursts could blow away any interstellar plasma surrounding the BH if $L_B$ exceed the BH’s Eddington luminosity. However, the brightest bursts generate a $e^-e^+$ plasma by Schwinger pair production, which is dense enough to prohibit photon propagation and block axion decay. Lasing may stop after a single laser pulse and restart once the plasma becomes subcritical again (via $e^-e^+$ annihilations), causing repeating bursts. Local plasma density or temperature, may also temporarily block lasing. By angular momentum conservation, one of the photons from each decay satisfies the superradiance condition $\omega_R<m\Omega_\text{BH}$, while the other is in a nonsuperradiant state, which may decrease the laser luminosity and modify its polarization through the spin-helicity effect \cite{Rosa:2015hoa,Rosa:2016bli,Leite:2017zyb}.

The parameters (mass and spin) of an axion cloud have to be smaller than those of its associated BH. This restricts the masses of axion and BH for $\alpha_\mu\lesssim0.05$ to be
$\mu\gtrsim10^{-8}\text{ eV}$, $M_\text{BH}\lesssim10^{-2}M_\odot$.
The notion that lasing can only occur in the non-relativistic regime is confirmed by $\alpha_\mu\lesssim0.03K$, which is obtained by requiring that quartic axion self-interactions are not significant. BLASTs can only occur for spinning primordial BHs(PBHs) \cite{Carr:1974nx} from merger of two nonspinning PBHs. 

The predicted peak luminosities and durations of the brightest BLASTs exhibits features similar to the fast radio bursts (FRBs) \cite{Lorimer:2007qn,Thornton:2013iua,Petroff:2016tcr} observed in recent years. A particular example is  the repeating FRB 121102 \cite{Spitler:2016dmz,Chatterjee:2017dqg,Marcote:2017wan,Tendulkar:2017vuq}. Geometrical and curved spacetime effects could prevent the observation of bursts along our line of sight and possibly explain why some FRBs do not repeat. 
The brightest BLASTs are expected to yield up to $\sim10^5$ FRBs per day across the whole sky. Lighter PBHs may become continuous laser sources ( since Schwinger pair production is less likely to occur in these cases) with lower luminosities, but they should be less numerous than repeating BLASTs due to their shorter lifetime \cite{Rosa:2017ury}. A similar comparison between the photon signature from superradiant pion clouds and the isotropic gamma-ray background (IGRB) is presented in  \cite{Ferraz:2020zgi}.

\subsection{Non-spherical cluster and static spacetime modification}
The evolution equations \eqref{EvoEq02} and \eqref{EvoEq04} are based on spherical symmetry $f_a(\vec{p})= f_a(|\vec{p}|)$, $f_\lambda(\vec{k})= f_\lambda(|\vec{k}|)$. One considers non-spherically symmetric  occupation numbers by allowing dependence on the directions of momenta $\vec{p}$ and $\vec{k}$. Let us write occupation numbers $f_a(\vec{p})= f_a(p, \Omega_p)$ and $f_\lambda(\vec{k})= f_a(p, \Omega_k)$ as spherical harmonic expansions,
\begin{flalign}  \label{}
f_a( \vec{p} ) = \sum_{lm}a_{lm}(p,t)Y_{lm}(\Omega_p) ~,~~
f_\lambda( \vec{k} ) = \sum_{lm}b_{lm}(k,t)Y_{lm}(\Omega_k) ~,
\end{flalign}
where $\Omega_p$ and $\Omega_k$ are the angular dependences from the directions of momenta $\vec{p}$ and $\vec{k}$, respectively. The evolution equations for individual components $b_{lm}$ of photon occupation number are given by equation (8) in \cite{Chen:2020ufn}, which is analogous to equation \eqref{EvoEq02}, without imposing spherical symmetry in momentum space.

Many of the equations cited here and elsewhere in this subsection from \cite{Chen:2020ufn,Chen:2020yvx,Chen:2023bne,Chen:2020eer} are and so not reproduced here. We simply provide a guide to them and refer the reader to those works for the details.

The corresponding rate equations for individual components of particle number densities are also obtained as equations (13--15) in \cite{Chen:2020ufn}.
Similarly, one relaxes spherical symmetry in coordinate/spatial space of simple axion cluster model \eqref{sss} by expanding occupation numbers $f_a(p, r, \Omega, t)$ and $f_\lambda(k, r, \Omega, t)$ in spherical harmonics \cite{Chen:2020yvx},
\begin{flalign}\nonumber
f_a(p, r, \Omega, t) =& \sum_{lm}f_{a lm}(t)Y_{lm}(\Omega)  \Theta(p_{\mbox{\tiny max}}-p)\, \Theta(R-r)  ~,  \\ 
 f_\lambda(k, r, \Omega, t) =& \sum_{lm}f_{\lambda lm}(t)Y_{lm}(\Omega) 
\Theta(R-r) \Theta(k_+-k) \Theta(k-k_-)  ~,
\end{flalign}
where $\Omega$ is the conventional spatial solid angle. The corresponding rate equations for individual components of particle number densities are then obtained as equations (19--21) in \cite{Chen:2020yvx}, which are later applied to a study \cite{Chen:2023bne} of the superradiant growth of non-spherical axions, as an update of the earlier analysis \cite{Rosa:2017ury}.

An axion cloud bounded by a host object in static spacetimes is studied and the rate equations of axion and photon number densities are given as equations (8.1--8.4) in \cite{Chen:2020eer}. Compare to the simple axion cluster model in Minkowski spacetime \cite{Kephart:1994uy}, an example \cite{Chen:2020eer} shows that the peak number density of photon is about 10\% larger and slightly delayed (see figure 4 in \cite{Chen:2020eer}), for a QCD axion ($m_a=10^{-5}$ eV) cluster of mass $M=6\times10^{21}$ kg a few meters away from the center of a Schwarzschild spacetime produced by a black hole of mass $M_\text{BH}=8\times10^{23}$ kg. Climbing out of the Schwarzschild
potential well, the associated tidal effects (gravitational redshift) initially slow the decay process but this wanes quickly and gives a sharper peak signal strength at a delayed time.

\section{Comments}
We have reviewed two approaches of investigations into the phenomenon of axion lasing, i.e., the field theory methods in Sec. II and the particle perspective in Sec. III.  
The results of the two methods are in qualitative agreement numerically and in general agreement in their conclusions.

There are many potential sources of astrophysical axion lasing. Depending on the axion mass and self coupling, these include clumps of axions and axion mini clusters and are limited by the objects angular momentum. In addition, lasing superradiant clouds of axions around primordial Kerr black holes provides an interesting possibility of discovery when identified with fast radio bursts. If verified this would constitute the discovery of two components of the dark matter, primordial  black holes and axions. 

More generally, axions and axion like particles are a compelling class of particles to investigate. They provide numerous and varied possibilities for physics, astrophysics and cosmological phenomena. They can arise in extensions of the standard model, grand unified theories and string theory. Their discovery would provide considerable insight into aspects of beyond standard model physics.

\begin{acknowledgments}
This work is supported in part by the National Key Research and Development Program of China under Grant No. 2020YFC2201501 and  the National Natural Science Foundation of China (NSFC) under Grant No. 12147103.
\end{acknowledgments}

\end{document}